\documentclass[conference,a4paper]{IEEEtran}
\IEEEoverridecommandlockouts

\usepackage{cite}
\usepackage{amsmath,amssymb,amsfonts}
\usepackage{algorithmic}
\usepackage{graphicx}
\usepackage{textcomp}
\usepackage{xcolor}
\usepackage{balance}

\newcommand{\argmax}{\mathop{\mathrm{argmax}}}

\usepackage{amsthm}
    \theoremstyle{definition}
    \newtheorem{example}{Example}
    \newtheorem{theorem}{Theorem}
\usepackage{cancel}
\usepackage{pgfplots}
\usepackage{tikz}

\definecolor{my_green}{rgb}{0.25, 0.6, 0.0}   
\pgfplotsset{compat=1.17} 

\def\BibTeX{{\rm B\kern-.05em{\sc i\kern-.025em b}\kern-.08em
    T\kern-.1667em\lower.7ex\hbox{E}\kern-.125emX}}
\begin{document}
\bstctlcite{IEEEexample:BSTcontrol}

\title{On the Design and Performance of Machine Learning Based Error Correcting Decoders}
\author{
    \IEEEauthorblockN{Yuncheng Yuan, Péter Scheepers, Lydia Tasiou, Yunus Can Gültekin, Federico Corradi, Alex Alvarado}

\thanks{The authors are with the Department of Electrical Engineering, Eindhoven University of Technology, 5600 MB Eindhoven, The Netherlands (corresponding author e-mail: a.alvarado@tue.nl).}

\thanks{This work has been partially funded by the Eurotech PhD program via the project CORTECH-4-PUF and by the Eindhoven Hendrik Casimir Institute (EHCI) Collaborative Projects Program.}
}

\maketitle

\begin{abstract}
This paper analyzes the design and competitiveness of four neural network (NN) architectures recently proposed as decoders for forward error correction (FEC) codes. 
We first consider the so-called single-label neural network (SLNN) and the multi-label neural network (MLNN) decoders which have been reported to achieve near maximum likelihood (ML) performance. 
Here, we show analytically that SLNN and MLNN decoders can \textit{always} achieve ML performance, regardless of the code dimensions---although at the cost of computational complexity---and no training is in fact required.
We then turn our attention to two transformer-based decoders: the error correction code transformer (ECCT) and the cross-attention message passing transformer (CrossMPT).
We compare their performance against traditional decoders, and show that ordered statistics decoding outperforms these transformer-based decoders. 
The results in this paper cast serious doubts on the application of NN-based FEC decoders in the short and medium block length regime.
\end{abstract}

\begin{IEEEkeywords}
Forward Error Correction, Machine Learning, Maximum Likelihood, Transformers.
\end{IEEEkeywords}

\section{Introduction}\label{introduction}
\IEEEPARstart{F}{orward} error correction codes are critical for reliable data transmission over noisy channels \cite{intro_surveyfec}. Enhancing FEC decoders to achieve near-ML decoding performance with low complexity and latency has been the holy grail of coding theory for many years. ML decoders minimize the probability of decoding errors and thus, they are performance-wise optimal.
However, their complexity grows exponentially with increasing code block length, making them impractical~\cite{ml_problem}.

Traditional (suboptimal) FEC decoders are generally categorized into the classes of hard- and soft-decision decoders.
Hard-decision decoders, e.g., the classic bit-flipping algorithms introduced by Gallager in \cite{bitflip} or bounded distance decoding for BCH codes, operate on binary data. 
While these methods are computationally simple, this simplicity often comes at the expense of relatively poor performance. 
On the other hand, soft-decision decoders, like the soft-input ML decoder or other near-ML decoders such as belief propagation (BP) for low-density parity-check codes, utilize probabilistic information about the received bits. Soft-decision decoders offer improved error correction performance at the cost of significantly higher complexity.

BP is an iterative decoding method for FEC codes, updating its estimates of the code bits via message passing \cite{1363033}. While very popular in practice, its complexity increases with the number of iterations and the block length. Another well-known soft-decision decoding method is ordered statistics decoding (OSD) \cite{fossorier}, which leverages soft-decision information by focusing on the most reliable code bit positions. Though effective, both BP and OSD face challenges in terms of complexity and latency, particularly for large block lengths.

In \cite{dlbasedchanneldecoding}, the concept of NN-based decoders is revisited, highlighting their potential for one-shot decoding and parallelizable architecture. 
Their results show that for the Polar~$(16,8)$~code the maximum a posteriori (MAP) performance is achieved, however, scaling to longer codes is prohibitive, due to exponentially increasing training complexity. 
Moreover, NN-based decoders, such as SLNN \cite{SLNN1, SLNN2, SLNN3} and MLNN \cite{SLNN1, SLNN3, MLNN1, MLNN2, MLNN3, MLNN4}, have emerged as potential efficient alternatives to {traditional} decoders, achieving ML or near-ML performance without performing exhaustive searches \cite{b9}. These model-free decoders \cite{dl-survey} do not rely on the knowledge of the specific code structure and are non-iterative, making them interesting in practice. 
However, the decoders from \cite{SLNN1, SLNN2, SLNN3, MLNN1, MLNN2, MLNN3, MLNN4} have not been extensively analyzed for different code types. Furthermore, their complexity compared to traditional decoders and their scalability to longer codes still remain unknown. 

Transformer-based decoders, including the error correction code transformer (ECCT) and the cross-attention message passing transformer (CrossMPT), have also recently been proposed \cite{ECCT}, \cite{park2024crossmptcrossattentionmessagepassingtransformer}. These decoders have shown improved decoding performance with respect to BP and BP-based neural decoders \cite{nbp1}.
While the results reported in \cite{ECCT, park2024crossmptcrossattentionmessagepassingtransformer} are promising, no comparisons exist against previously proposed NN-based decoders such as SLNN and MLNN. In addition, apart from some simple comparisons in \cite{ECCT,park2024crossmptcrossattentionmessagepassingtransformer}, no extensive comparisons against traditional decoders exist in the literature.

In this paper, we first show analytically that SLNN and MLNN decoders achieve exact ML performance for a specific network architecture defined by the FEC code structure, without even any training. This is achieved by using weights that are simply the codewords in the codebook. 
We then expose that these network architectures are significantly less complex than the ones used in \cite{SLNN2, SLNN3, MLNN1} to achieve near-ML performance.
Next, we reproduce performance results of the ECCT and CrossMPT decoders from \cite{ECCT, park2024crossmptcrossattentionmessagepassingtransformer} and compare them, for the first time, to OSD. We show that these transformer-based decoders are not competitive to OSD.

The organization of the paper is as follows. 
Preliminaries are introduced in Section~\ref{methods}. 
SLNN and MLNN decoders are investigated in Sec.~\ref{MLdecoder}. 
The performance and limitations of ECCT andCrossMPT are analyzed in Sec.~\ref{Tdecoder}. 
Conclusions are future prospects are given in Sec.~\ref{conclusion}.

\section{Preliminaries}\label{methods}
\subsection{System Model and Maximum Likelihood Decoding}

The communication system under consideration includes three main components: the transmitter, the channel, and the receiver. The transmitter is a concatenation of an FEC encoder and a binary phase-shift keying (BPSK) mapper. The receiver consists of a soft-input FEC~decoder. At the transmitter, the information bits $\boldsymbol{b} = (b_1, \dots, b_k)$ are encoded using an $(n,k)$ linear block code into a codeword $\boldsymbol{c} = (c_1, \dots, c_n)$. 
Encoding is achieved through $\boldsymbol{c} = \boldsymbol{b} \boldsymbol{G}$, where $\boldsymbol{G}$ is the $k \times n$ generator matrix of the code. The codebook $\mathcal{C} = \{ \boldsymbol{c}^{(1)}, \dots, \boldsymbol{c}^{(2^k)}\}$ consists of the $2^k$ possible codewords. 

After encoding, the code bits $c_i$ are mapped to BPSK symbols $s_i \in \{-1, 1\}$ via $s_i = 2c_i -1$, for $i = 1, \dots, n$. The transmitted symbols $\boldsymbol{s} = (s_1, \dots, s_n)$ are sent over a binary-input additive white Gaussian noise (BI-AWGN) channel. At the receiver, the real-valued noisy symbols $\boldsymbol{r} = \boldsymbol{s} + \boldsymbol{w}$ are observed, where $\boldsymbol{w} = (w_1, \dots, w_n)$ is an $n$-dimensional vector of independent and identically distributed (i.i.d.) noise samples $w_i$, drawn from a zero-mean Gaussian distribution with variance $\sigma^2 = {N_0}/{2}$ where $N_0$ is the power spectral density of noise. 
Throughout this paper, we report results as a function of ${E_b}/{N_0}$, the so-called ``bit SNR'', where the energy per information bit $E_b=1/(k/n)$ assuming $s_i=\pm 1$ are equally-likely. 
At the receiver, the transmitted codeword $\hat{\boldsymbol{c}}$ is estimated by applying a decoding algorithm to the received noisy symbols $\boldsymbol{r}$.

For any $(n,k)$ linear block code whose codewords are transmitted with equal probability, the ML decoding rule is 
\begin{equation}
     \hat{\boldsymbol{c}} = \argmax_{\boldsymbol{c} \in \mathcal{C}} p(\boldsymbol{r} | \boldsymbol{s}),
     \label{eq:mlcodeword}
\end{equation}
where $p(\boldsymbol{r} | \boldsymbol{s})$ is the likelihood of receiving $\boldsymbol{r}$ when $\boldsymbol{c}$ was transmitted, and $\hat{\boldsymbol{c}}$ is the estimated codeword. 
For a memoryless channel, \eqref{eq:mlcodeword} can be rewritten as
\begin{align}
    \hat{\boldsymbol{c}} &= \argmax_{\boldsymbol{c} \in \mathcal{C}} \sum_{i=1}^{n} \log(p(r_i | s_i)).
    \label{eq:sumlogp}
\end{align}
For the BI-AWGN channel, the channel transition probability density function $p(r_i|s_i)$ is given by
\begin{equation}
    p(r_i|s_i) = \frac{1}{\sqrt{2 \pi \sigma^2}} \textrm{exp}\left( - \frac{(r_i - s_i)^2}{2\sigma^2}\right).
    \label{eq:awgnexp}
\end{equation}
Substituting (\ref{eq:awgnexp}) into (\ref{eq:sumlogp}) gives
\begin{align}
    \hat{\boldsymbol{c}} &= \argmax_{\boldsymbol{c} \in \mathcal{C}} \left( -\frac{n}{2} \log{(2 \pi \sigma^2 )} - \frac{1}{2\sigma^2} \sum_{i=1}^{n} (r_i - s_i)^2 \right), \label{eq:3in2}
\end{align}
where the first term can be neglected as it does not depend on~$\boldsymbol{c}$, i.e., on $\boldsymbol{s}$. 
Expanding the second term in \eqref{eq:3in2} results in 
\begin{equation}
    \hat{\boldsymbol{c}} = \argmax_{\boldsymbol{c} \in \mathcal{C}} - \frac{1}{2\sigma^2} \left( ||\boldsymbol{r}||_2^2 - \sum_{i=1}^{n} 2 r_i s_i + ||\boldsymbol{s}||_2^2 \right), \label{eq:expanded}
\end{equation}
where $||\cdot||_2^2$ is the squared $L_2$ norm. 
Since $||\boldsymbol{s}||_2^2$ is constant for all $\boldsymbol{c} \in \mathcal{C}$ and $||\boldsymbol{r}||_2^2$ does not depend on $\boldsymbol{c}$, they can both be ignored. 
Using the fact that $s_i = 2c_i - 1$ in \eqref{eq:expanded} and further neglecting the terms that do not depend on~$\boldsymbol{c}$,  we can express the ML rule for the BI-AWGN channel as
\begin{equation}
    \hat{\boldsymbol{c}} = \argmax_{\boldsymbol{c}^{(j)} \in \mathcal{C}} 
    \sum_{i=1}^n ( 2c_i^{(j)}-1) r_i = \argmax_{\boldsymbol{c}^{(j)} \in \mathcal{C}} 
    \sum_{i=1}^n c_i^{(j)} r_i,  
    \label{eq:mlcorr}
\end{equation}
for $j = 1,\dots, 2^k$ where $c_{i}^{(j)}$ is the $i^{\text{th}}$ bit of the $j^{\text{th}}$ codeword in the codebook $\mathcal{C}$, and $r_i$ is the $i^{\text{th}}$ received noisy symbol in $\boldsymbol{r}$.
The ML decoder estimates the code bits based on real-valued channel outputs, i.e., it is a soft-input hard-output decoder.

\subsection{Ordered Statistics Decoding}\label{OSD}

OSD is a soft-input decoding algorithm particularly well-suited for short and medium  blocklengths. OSD sorts the vector $\boldsymbol{r}$ according to the permutation $\Phi_1$ based on the reliability values as
    \begin{equation*}
        \Phi_1(\boldsymbol{r}) = \boldsymbol{r}' = (r'_1, \ldots, r'_n) \quad \mbox{such that} \quad |r'_1|\geq \ldots \geq |r'_n|.
    \end{equation*}
The permutation $\Phi_1$ is applied to the columns of $\boldsymbol{G}$, denoted by $\boldsymbol{G}' = \Phi_1(\boldsymbol{G})$, to obtain an equivalent matrix $\boldsymbol{G}'$ with the corresponding (equivalent) code $\mathcal{C'}$.
Then, the vector $\boldsymbol{r}'$ is permuted by a second permutation $\Phi_2$, which is based on the so-called ``most reliable basis''. This basis is determined as follows: starting from the first column of $\boldsymbol{G}'$, the first $k$ linearly independent columns with the highest corresponding reliability values are found. The corresponding reliability of the $i$-th column of $\boldsymbol{G}'$ refers to the value $|r'_i|$, $1\leq i \leq n$. 
    \begin{equation*}
        \boldsymbol{r}'' = (r''_1, \ldots, r''_n) = \Phi_2(\Phi_1(\boldsymbol{r})),
    \end{equation*} 
with $|r''_1| \geq \ldots \geq |r''_k| \wedge  |r''_{k+1}| \geq \ldots \geq |r''_n|$. 
    
Using row operations, $\boldsymbol{G}''=\Phi_2(\Phi_1(\boldsymbol{G}))$ is brought into a systematic form. The first $k$ coordinates act as an information set. Hard-decisions based on $\boldsymbol{r}''_A = (r''_1, \ldots, r''_k)$ are then used to generate a codeword $\boldsymbol{r}_0$ in the equivalent code $\mathcal{C}''$. Lastly, 
reverse permutations are applied to obtain a codeword $\hat{\boldsymbol{c}}$ in the original code: $\hat{\boldsymbol{c}} = \Phi_2^{-1}(\Phi_1^{-1}(\boldsymbol{r}_0))$.
We refer to this method as OSD of order $q=0$. For $q>0$, test patterns are generated by adding to $\boldsymbol{r}''_A$ error patterns of Hamming weight up to $q$.

\section{Neural Network Decoders}\label{MLdecoder}
\subsection{Single-Label Neural Network Decoder}
\label{subsec:slnn}

\begin{figure}[tpb]
\centering
	\begin{minipage}[b]{0.24\textwidth}
		\resizebox{\textwidth}{!}{\includegraphics{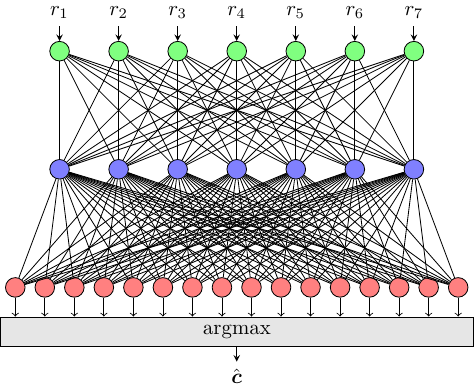}}
	\end{minipage}
	\begin{minipage}[b]{0.24\textwidth}
		\resizebox{\textwidth}{!}{\includegraphics{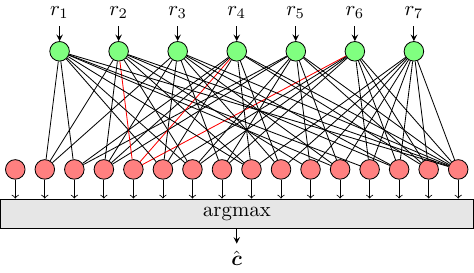}}
	\end{minipage}
\caption{(Left) SLNN decoder architecture SLNN $7$-$7$-$16$ in~\cite{SLNN2, SLNN3} for the Hamming~$(7,4)$ code. (Right) SLNN architecture SLNN $7$-$16$ proposed in this work (see Theorem~\ref{theorem1}), where the hidden layer is not present. The red edges in the right figure represent the codeword used in Example~\ref{ECCT Hamming(7,4)}.}
\vspace{-2ex}
\label{fig:7-4-no-hidden-layer}
\end{figure}

In~\cite{SLNN1, SLNN2, SLNN3}, the decoding problem was formulated as a supervised single-label classification problem. 
The proposed SLNN decoder is a fully-connected network with $N_0 = n$ input neurons, one hidden layer with $N_1$ neurons applying ReLU activation, and $N_{2} = 2^k$ output neurons using scaled softmax, with scaling factor $\alpha$ equal to 1, to output a probability distribution over $2^k$ possible codewords. 
The scaled softmax function is given by
\begin{equation}\label{eq:scaledsoftmax}
    \sigma(x_i) = \frac{\textrm{exp}\left(\alpha x_i \right)}{\sum_{j=1}^{M} \textrm{exp}(\alpha x_j)},
\end{equation}
where $x_i$ is the input to the $i$-th neuron, and $M$ is the total number of neurons in the layer. 
The estimated transmitted codeword, i.e., the output neuron with the highest probability, is found by applying the $\argmax$ operation, see Fig.~\ref{fig:7-4-no-hidden-layer} (left).

In this paper, the SLNN decoder is trained following the approach in \cite{SLNN2}, i.e., based on a dataset $\mathcal{D}_{\text{SLNN}}$ that consists of pairs of $\boldsymbol{r}^{(j)} \in \mathbb{R}^n$ and $\boldsymbol{x}^{(j)} \in \{ 0, 1\}^{2^k}$ for $j = 1, 2,\dotsc, |\mathcal{D}_{\text{SLNN}}|$.
Here, $\boldsymbol{r}^{(j)}$ is the $j^{\text{th}}$ noisy received codeword in the dataset
and $\boldsymbol{x}^{(j)}$ is the $j^{\text{th}}$ one-hot encoded label corresponding to the codeword in $\mathcal{C}$ that led to the reception of $\boldsymbol{r}^{(j)}$.
The dataset $\mathcal{D}_{\text{SLNN}}$ is generated via uniform random sampling from the codebook and transmission at $0$ dB SNR, which has been shown in \cite{SLNN2} to generalize well across SNRs and achieve near-ML performance. 
The dataset is split into training and test (validation) sets.
The accuracy is measured via the fraction of codewords correctly classified over the total number of codewords. 
Training continues until the complement of the accuracy, i.e. the error rate, is equal to the frame error rate (FER) found at $0$ dB~SNR using the ML decoder. 

\begin{figure}[!t]
    \centering
   \resizebox{0.5\columnwidth}{!}{\includegraphics{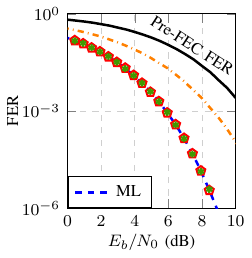}}
    \hspace{-2ex}
    \resizebox{0.5\columnwidth}{!}{\includegraphics{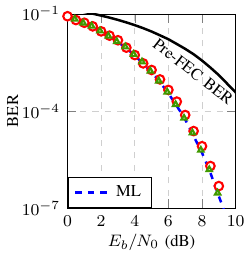}}
        
    \resizebox{0.8\columnwidth}{!}{\begin{tikzpicture}
	\begin{axis}[
		width=1.1\textwidth,
		hide axis,
		xmin=0,
        xmax=0,
        ymin=0,
        ymax=0,
        legend style={font=\tiny,row sep=-0.5ex,inner sep=0.1ex},
        legend columns = 2,
         legend cell align={left}
		]

        \addlegendimage{mark=*, mark size = 1pt, dashdotted, color=orange, no marks, line width=0.75pt,  mark options={solid, purple, scale=.75, fill=white}}
        \addlegendentry{SLNN $7$-$5$-$16$ \cite{SLNN2, SLNN3}}

        \addlegendimage{mark=*, mark size = 2pt, color=red, line width=0.75pt,  only marks, mark options={solid, red, scale=.75, fill=white}}
        \addlegendentry{MLNN $7$-$50$-$50$-$4$ \cite{SLNN3, MLNN1}}

        \addlegendimage{mark=pentagon*, mark size = 2pt, color=red, line width=0.75pt,  only marks, mark options={solid, red, scale=.75, fill=white}}
        \addlegendentry{SLNN $7$-$7$-$16$ \cite{SLNN2, SLNN3}}

        \addlegendimage{mark=triangle*, mark size = 2pt, color=my_green, line width=0.75pt, only marks, mark options={solid, my_green, scale=.75, fill=white}}
        \addlegendentry{MLNN $7$-$16$-$4$ (Theorem~\ref{theorem2})\phantom{xx}}
        
        \addlegendimage{mark=asterisk, mark size = 2pt, color=my_green, line width=0.75pt, only marks, mark options={solid, my_green, scale=.75, fill=white}}
        \addlegendentry{SLNN $7$-$16$ (Theorem~\ref{theorem1})}

	\end{axis}  
\end{tikzpicture}}
    \caption{FER (left) and BER (right) vs. SNR for different SLNN (left) and MLNN (right) decoders for the Hamming~$(7,4)$ code. The results for SLNN $7$-$5$-$16$, SLNN $7$-$7$-$16$, and  MLNN $7$-$50$-$50$-$4$ are based on our implementation of the architectures proposed in \cite{SLNN2, SLNN3, MLNN1}.}
    \vspace{-2ex}
\label{fig:MLNN_hamm74_84_ber}
\end{figure}

\begin{example}[SLNN for Hamming~$(7,4)$]
The SLNN decoder architecture was implemented using PyTorch to train the SLNN decoder and to simulate the decoding performance via Monte Carlo simulations. An example of this architecture is shown in Fig.~\ref{fig:7-4-no-hidden-layer}~(left) for the simple case of a Hamming~$(7,4)$ code. The notation used here for SLNN indicates the number of neurons in each layer, e.g., SLNN $7$-$7$-$16$ means an input layer with $N_0=7$ neurons, a hidden layer with $N_1=7$ neurons, and an output layer with $N_2=16$ neurons. FER results are shown in Fig.~\ref{fig:MLNN_hamm74_84_ber} (left). For each SNR point, simulations were performed until at least $2000$ frame errors were detected. These results show that SLNN with a single hidden layer can achieve ML performance, as long as enough neurons in the hidden layer are used. For the Hamming~$(7,4)$ code, $N_1=5$ is not enough. As shown in \cite{SLNN2, SLNN3}, when $N_1=n$ ($N_1=7$ here), the ML performance is achieved.
\end{example}

The following theorem is the first contribution of this paper, which shows that the hidden layer in the SLNN architecture is, in fact, not needed. ML performance can be always guaranteed for any $(n,k)$ linear block code with an NN with $N_0=n$ input neurons and $N_2=2^k$ output neurons.

\begin{theorem}
    \label{theorem1}
    Let $(n,k)$ be a linear block code, with codebook $\mathcal{C}$ containing $2^k$ codewords. Consider a two-layer NN with $n$ input neurons and $2^k$ output neurons. Let the $n$-by-$2^k$ weight matrix $\boldsymbol{W}^{(1)}$ connecting the input layer to the output layer be binary, and has its columns equal to the codewords in $\mathcal{C}$. This NN architecture realizes (codeword-wise) ML decoding. No training is required.
\end{theorem}
\begin{IEEEproof}
The summation in \eqref{eq:mlcorr} can be expressed as a vector multiplication between $\boldsymbol{c}_\mathrm{T}^{(j)}$ and $\boldsymbol{r}$, where $(\cdot)_\mathrm{T}$ indicates the transpose operation, i.e., $ \hat{\boldsymbol{c}} = \argmax_{\boldsymbol{c}^{(j)} \in \mathcal{C}} \boldsymbol{r} \boldsymbol{c}^{(j)}_\mathrm{T} $.
The proof is completed by representing this maximization problem in matrix form: Consider a vector-matrix multiplication between the received signal $\boldsymbol{r}$ and the $n$-by-$2^k$ matrix $\boldsymbol{W}^{(1)}$ for which each column is equal to a different $\boldsymbol{c} \in \mathcal{C}$. 
The $j^{\text{th}}$ element in the vector $\boldsymbol{r}\boldsymbol{W}^{(1)}$ is equal to the $j^{\text{th}}$ metric $\boldsymbol{r}\boldsymbol{c}^{(j)}_\mathrm{T}$ over which the maximization is carried out.
This is what the SLNN architecture with $N_0=n$, $N_1=0$, and $N_2=2^k$ realizes.
\end{IEEEproof}

The result in Theorem~\ref{theorem1} shows that the SLNN decoder without the hidden layer realizes ML decoding. 
Furthermore, this result also implies that no training is required, as the weights of the NN are in fact the codewords of the code to be decoded. The number of edges in the resulting NN is the sum of the Hamming weights of all codewords in $\mathcal{C}$.

\setcounter{example}{0}

\begin{example}[SLNN for Hamming~$(7,4)$ (Cont.)]
Fig.~\ref{fig:7-4-no-hidden-layer} (right) shows the SLNN architecture based on Theorem~\ref{theorem1} for the Hamming~$(7,4)$ code. This figure shows that there is no hidden layer, and the input neurons are sparsely connected to the output neurons through the elements of the codebook. The left-most neuron in the output layer is in fact not connected at all, which is due to the fact that the first codeword in the codebook is the all-zero codeword.
Fig.~\ref{fig:MLNN_hamm74_84_ber} (left) shows the FER results for SLNN $7$-$16$. As expected, the presented results show that the hidden layer in SLNN $7$-$7$-$16$ is indeed redundant, and only increases the computational complexity and memory requirements of the decoder. For this code, the number of edges in the NN is lowered from $161$ (for SLNN $7$-$7$-$16$) to only $56$ (for SLNN $7$-$16$).
\end{example}

\subsection{Multi-Label Neural Network Decoder}
\label{subsec:mlnn}

\def\boldc{\boldsymbol{c}}
\def\boldr{\boldsymbol{r}}

\begin{figure}[tpb]
\centering
\begin{minipage}[b]{0.24\textwidth}
\resizebox{\textwidth}{!}{\includegraphics{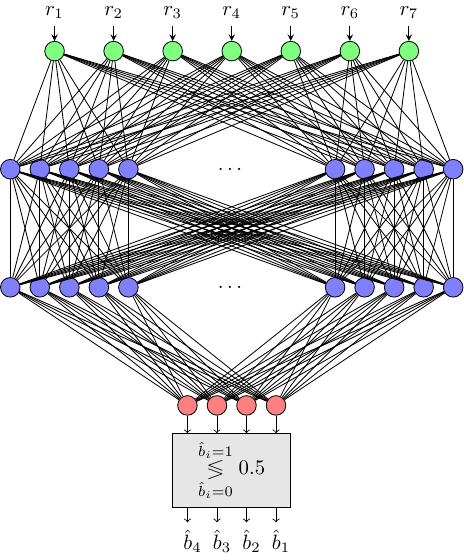}}
\end{minipage}
\begin{minipage}[b]{0.24\textwidth}
\resizebox{\textwidth}{!}{\includegraphics{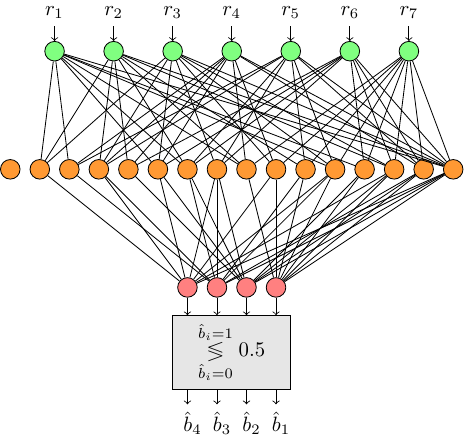}}
\end{minipage}
\caption{(Left) MLNN decoder $7$-$50$-$50$-$4$ in~\cite{SLNN3, MLNN1} for the Hamming~$(7,4)$ code. (Right) MLNN decoder $7$-$16$-$4$ proposed in this~work.}
\vspace{-2ex}
\label{fig:mlnn_7_4_comparison}
\end{figure}

In \cite{SLNN1, SLNN3, MLNN1, MLNN2, MLNN3, MLNN4}, the decoding problem was cast as a supervised multi-label classification problem. 
The proposed MLNN architecture is a fully-connected NN with $N_0 = n$ input neurons and $N_{L+1} = k$ output neurons, where a sigmoid activation function is applied to estimate the probability of each bit.
The number of hidden layers $L$ and hidden layer neurons $N_i$ for $i=1, 2,\dotsc, L$ are design parameters.
The activation function in the hidden layers is ReLU. 

To train the MLNN decoder, we follow the approach in \cite{MLNN1}, where a dataset $\mathcal{D}_{\text{MLNN}}$ consisting of pairs $\boldsymbol{r}^{(j)} \in \mathbb{R}^n$ and $\boldsymbol{b}^{(j)} \in \{0, 1\}^k$ for $j=1,2,\dots,|\mathcal{D}_{\text{MLNN}}|$ is generated. 
$\boldsymbol{r}^{(j)}$ is the $j^{\text{th}}$ noisy received codeword and $\boldsymbol{b}^{(j)}$ is the $j^{\text{th}}$ information bit vector that led to the reception of $\boldsymbol{r}^{(j)}$. 
The dataset $\mathcal{D}_{\text{MLNN}}$ is generated via uniform random sampling from all possible information bit vectors and transmission at $0$ dB SNR, which leads to good generalization across SNRs and achieves near-ML performance \cite{MLNN1}. 
Here, the accuracy is the fraction of correctly classified information bits over the total number of information bits. 
Training continues until the complement of the accuracy is equal to the bit error rate (BER) found at $0$ dB~SNR using the ML decoder. 

\begin{example}[MLNN for Hamming~$(7,4)$]
In \cite{SLNN3, MLNN1}, it was found that for the Hamming~$(7,4)$ code, $2$ hidden layers with $50$ neurons each are sufficient to achieve near-ML performance. Fig.~\ref{fig:mlnn_7_4_comparison} (left) shows the resulting architecture, where we use a notation analogous to that used in Sec.~\ref{subsec:slnn}.
BER results are shown in Fig.~\ref{fig:MLNN_hamm74_84_ber} (right). 
For each SNR point, simulation was run until $2000$ bit errors were detected. These results show that near-ML performance is indeed achieved using MLNN $7$-$50$-$50$-$4$. This performance is achieved, however, using an NN with $3200$ edges, see Fig.~\ref{fig:mlnn_7_4_comparison} (left).
\end{example}

The next theorem is the second contribution of this paper, which shows that there exists an MLNN architecture with a single hidden layer that does not have to be trained and achieves ML performance. The number of edges of such an NN is also much lower than that required in \cite{SLNN3, MLNN1}.

\begin{theorem}
    \label{theorem2}
    Let $(n,k)$ be a linear block code $\mathcal{C}$. Consider a three-layer NN with $n$ input neurons, $2^k$ hidden neurons applying the scaled softmax in \eqref{eq:scaledsoftmax} with $\alpha=2/\sigma^2$ as activation function, and $k$ output neurons. 
    Let the $n$-by-$2^k$ weight matrix $\boldsymbol{W}^{(1)}$ connecting the input layer to the hidden layer be binary, and has its columns equal to the codewords in $\mathcal{C}$.
    Let the $2^k$-by-$k$ weight matrix $\boldsymbol{W}^{(2)}$ connecting the hidden layer to the output layer be binary, and has its rows equal to the all possible $k$-bit input strings of the code.
    We assume that the columns and rows of $\boldsymbol{W}^{(1)}$ and $\boldsymbol{W}^{(2)}$, resp., are sorted such that $j^{\text{th}}$ row of $\boldsymbol{W}^{(2)}$ is encoded into $j^{\text{th}}$ column of $\boldsymbol{W}^{(1)}$ for $j = 1, 2,\dotsc, 2^k$.
    This NN architecture realizes (bit-wise) ML decoding.
    No training is~required.
\end{theorem}

\begin{IEEEproof}
The bit-wise maximum a posteriori (MAP) decoding rule is given by
\begin{equation}
    \hat{c}_{i} = \argmax_{b \in \{0, 1\}} p(c_i=b | \boldsymbol{r}),
    \label{eq:mapcri}
\end{equation}
for $i = 1, 2,\cdots, n$, which is equivalent to
\begin{equation}\label{eq:compareMAP}
    p(c_i = 0|\boldsymbol{r}) \underset{\hat{c}_i = 0}{\overset{\hat{c}_i = 1}{\lessgtr}} p(c_i = 1|\boldsymbol{r}),
\end{equation}
which can be rewritten using Bayes' rule as
\begin{equation}\label{eq:Bayesequiv}
 \sum_{\boldc: c_i=0} \frac{p(\boldr|\boldc) p(c_i=0)}{p(\boldr)} \underset{\hat{c}_i = 0}{\overset{\hat{c}_i = 1}{\lessgtr}} \sum_{\boldc: c_i=1} \frac{p(\boldr|\boldc) p(c_i=1)}{p(\boldr)}.
\end{equation}
Cancelling out $p(c_i=0)=p(c_i=1)$---which holds for linear codes with independent and uniform input bits---and $p(\boldr)$ on both sides, the left-hand side of \eqref{eq:Bayesequiv} can be written as
\begin{equation}
    \sum_{\boldc: c_i=0} \prod_{j=1}^{n} \frac{1}{\sqrt{2\pi\sigma^2}} \exp\left( -\frac{(r_j-(2c_j-1))^2}{2\sigma^2} \right),
\end{equation}
which can be expanded, ignoring ${1}/{\sqrt{2\pi\sigma^2}}$, into
\begin{equation*}
    \sum_{\boldc: c_i=0} 
    \exp\left(-\sum_{j=1}^{n} \frac{\cancel{r_j^2}-4r_jc_j +\cancel{2r_j}+\cancel{4c_j^2}-\cancel{4c_j}+\cancel{1}}{2\sigma^2}\right).
\end{equation*}
Neglecting common terms (crossed out) in \eqref{eq:Bayesequiv} yields
\begin{equation}
    \sum_{\boldc: c_i=0} \exp\left( \frac{2\boldsymbol{r}\boldsymbol{c}_\mathrm{T} }{\sigma^2}  \right) \underset{\hat{c}_i = 0}{\overset{\hat{c}_i = 1}{\lessgtr}} \sum_{\boldc: c_i=1} \exp\left( \frac{2 \boldsymbol{r}\boldsymbol{c}_\mathrm{T}}{\sigma^2}  \right).
\end{equation}
Dividing both sides by the sum of both sides, we get
\begin{equation}
    \sum_{\substack{\boldsymbol{c}: c_i = 0}}  \frac{\exp\left( \frac{2 \boldsymbol{r}\boldsymbol{c}_\mathrm{T}}{\sigma^2} \right)}{\sum_{\boldc} \exp\left( \frac{2\boldsymbol{r}\boldsymbol{c}_\mathrm{T}}{\sigma^2} \right)} \underset{\hat{c}_i = 0}{\overset{\hat{c}_i = 1}{\lessgtr}}  \sum_{\substack{\boldsymbol{c}:c_i = 1}} \frac{\exp\left( \frac{2\boldsymbol{r}\boldsymbol{c}_\mathrm{T}}{\sigma^2} \right)}{\sum_{\boldc} \exp\left( \frac{2\boldsymbol{r}\boldsymbol{c}_\mathrm{T}}{\sigma^2} \right)}.
    \label{eq:comparisontheorem2}    
\end{equation}
In the NN we defined in Theorem~\ref{theorem2}, the input-to-hidden transformation with the scaled softmax activation \eqref{eq:scaledsoftmax} computes the addends of the outer summations on both sides of \eqref{eq:comparisontheorem2}.
Then the hidden-to-output transformation computes the outer summation.
Also observing that left- and right-hand sides of \eqref{eq:comparisontheorem2} add up to $1$, a thresholding operation applied on the output nodes of the NN, with threshold set at $1/2$, results in \eqref{eq:mapcri}. 
\end{IEEEproof}

\setcounter{example}{1}

\begin{example}[MLNN for Hamming~$(7,4)$ (Cont.)]
Fig.~\ref{fig:MLNN_hamm74_84_ber} (right) shows the results for the MLNN $7$-$16$-$4$ (Thm.~\ref{theorem2}) for the Hamming~$(7,4)$ code. 
The proposed MLNN decoders in the literature do not achieve ML results for increasing SNR, see, e.g.,~\cite[Figs. 5 \& 8]{SLNN3}. 
In contrast, the MLNN architecture based on Thm. \ref{theorem2} realizes ML decoding for all SNRs with fewer hidden layer neurons and without training.
Another difference with the MLNN decoders proposed in the literature is that those use ReLU activation in the hidden layers, while the MLNN architecture from Thm. \ref{theorem2} uses softmax activation function. Although ReLU activation is computationally simpler than softmax, the overall dimensions of the MLNN decoders in the literature are larger, resulting in computationally more expensive matrix-vector multiplications. From Fig.~\ref{fig:mlnn_7_4_comparison}, it is clear that the MLNN architecture from the literature (left) is significantly larger ($3200$ edges) compared to the one proposed in Thm. \ref{theorem2} (right, with $88$ edges), while the latter realizes ML decoding.
Note that the softmax scaling factor $2/\sigma^2$ is SNR-dependent. 
\end{example}

\begin{figure*}[t]
    \centering    
    \resizebox{0.361\textwidth}{!}{\includegraphics{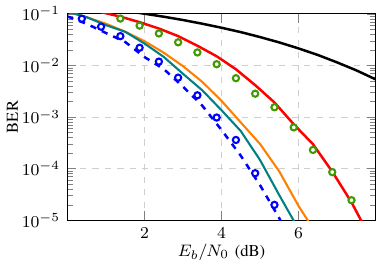}}
    \resizebox{0.3115\textwidth}{!}{\includegraphics{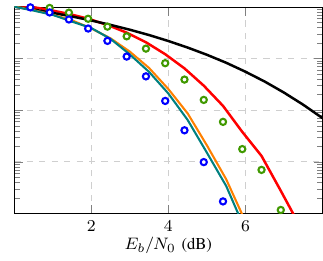}}
    \resizebox{0.3135\textwidth}{!}{\includegraphics{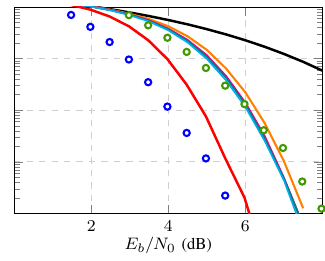}}
    \resizebox{\textwidth}{!}{\definecolor{my_green}{rgb}{0.25, 0.6, 0.0}   
\begin{tikzpicture}
	\begin{axis}[
		width=\textwidth,
        height=20ex,
		hide axis,
		xmin=0,
        xmax=0,
        ymin=0,
        ymax=0,
        legend style={font=\scriptsize},
        legend columns = 9,
        legend cell align={left}
		]
        \addlegendimage{mark=, color=black, solid, line width=1.2pt, forget plot}
        \addlegendentry{Pre-FEC\hphantom{x}}

        \addlegendimage{blue, thick, dashed, line width=1.0pt}
        \addlegendentry{Maximum Likelihood\hphantom{x}}
            
		\addlegendimage{color=red, solid, line width=1.0pt}
	    \addlegendentry{HD (BDD) Decoding\hphantom{x}}

        \addlegendimage{mark=*, mark size = 2pt, color=my_green, only marks, line width=1.0pt,mark options={solid, color=my_green, scale=.75, fill=white}}
        \addlegendentry{OSD $q=0$\hphantom{x}}

        \addlegendimage{mark=*, mark size = 2pt, color=blue, only marks, line width=1.0pt,mark options={solid, color=blue, scale=.75, fill=white}}
        \addlegendentry{OSD $q=1$\hphantom{x}}
        
	    
        \addlegendimage{color=orange, no marks, mark=triangle*, solid, line width=1.0pt, mark options={solid, orange, scale=.75, fill=white}}
        \addlegendentry{ECCT $N=6$\hphantom{x}}

        \addlegendimage{color=purple, no marks, mark=triangle*, solid, line width=1.0pt, mark options={solid, purple, scale=.75, fill=white}}
        \addlegendentry{ECCT $N=10$\hphantom{x}}

        \addlegendimage{color=teal, dashdotted, mark=x*, solid, line width=1.0pt, mark options={solid, teal, scale=.75, fill=white}}
        \addlegendentry{CrossMPT $N=6$\hphantom{x}}

        \addlegendimage{color=cyan, dashdotted, mark=x*, solid, line width=1.0pt, mark options={solid, cyan, scale=.75, fill=white}}
        \addlegendentry{CrossMPT $N=10$\hphantom{x}}
       
	\end{axis}  
\end{tikzpicture}
}
    \vspace{-0.45cm}
    \caption{BER vs. SNR for different BCH codes: $(31,16)$ (left), $(63,51)$ (middle), and $(127,64)$ (right). Both ECCT and CrossMPT results are obtained with $d=128$. For the $(31,16)$ BCH code (left), ML decoding is also shown. For the $(127,64)$ BCH code (right), the ECCT result with $N=10$ is also shown.}
    \vspace{-2ex}
    \label{fig:ECCT results}
\end{figure*}

\balance

\section{Transformer-based Decoders}\label{Tdecoder}
In this section we focus on recent proposals for using transformer architectures, originally proposed in \cite{vaswani2023attentionneed} for large language model applications, for error correcting codes.

\subsection{ECCT and CrossMPT Decoders}
In \cite{ECCT}, the decoding problem was formulated as a denoising task, with noise characteristics learned via the self-attention mechanism in transformers. ECCT utilizes the magnitude of the channel observations and the syndrome to learn the multiplicative noise, $\tilde{w}_{i}$, defined as
    $r_{i} = s_{i} + w_{i} = s_{i}\tilde{w}_{i}$. 
In what follows we briefly review the operation of ECCT. The reader is referred to \cite[Sec.~IV]{ECCT} for more details.

In the pre-processing step of ECCT, the received vector \( \boldsymbol{r} \) is transformed into a new vector \( \tilde{\boldsymbol{r}} \) of length \( n + (n-k) \), which consists of the concatenation of \( |r_1|,|r_2|,\ldots,|r_n| \) and the sign of the syndrome \( \boldsymbol{s}(\boldsymbol{r}) \). 
The sign of syndrome is defined as \( \boldsymbol{s}(\boldsymbol{r}) = 2 (\boldsymbol{H} \boldsymbol{r}_\text{hd} -1)\), where $\boldsymbol{r}_\text{hd}$ the vector of hard-decisions made on $\boldsymbol{r}$, and where \( \boldsymbol{H} \) is the parity-check matrix of the code. The vector $\boldsymbol{s}(\boldsymbol{r})$ therefore represents a BPSK-modulated version of the syndrome $\boldsymbol{H} \boldsymbol{r}_\text{hd}$.  

\begin{example}[ECCT for Hamming~$(7,4)$]
\label{ECCT Hamming(7,4)}
Consider the Hamming~$(7,4)$ code with transmitted codeword \(\boldsymbol{c}_{i}=[0, 1, 0, 0, 1, 1, 0] \) and received noisy codeword \( \boldsymbol{r} = [-0.75, 0.98, -0.31, 0.18, 3.08, 1.08, -1.23] \). The vector of hard-decisions is $\boldsymbol{r}_\text{hd}=[0,1,0,1,1,1,0]$
and the syndrome sign is \( \boldsymbol{s}(\boldsymbol{r}) = [1, 1, 1] \).
We then have $\tilde{\boldsymbol{r}} = [0.75, 0.98, 0.31, 0.18, 3.08, 1.08, 1.23, 1, 1, 1]$.
\end{example}

In the pre-processing step, the magnitudes $|r_i|$ assist the model in learning the multiplicative noise, while the syndrome sign identifies the presence of errors in the received noisy codeword, providing critical information to the transformer model. This additional feature helps the transformer focus its attention on the most likely noised parts, improving the overall denoising accuracy. Given that ECCT requires knowledge on the received vector $\boldsymbol{r}$, it is a soft-input decoding algorithm.

In the decoding process, the received noisy codeword is processed by a self-attention transformer. The model begins by converting the input vector \( \tilde{\boldsymbol{r}} \) into embeddings. The model reformats input data into a high-dimensional representation, helping the model capture complex patterns in the noisy codeword more effectively than with the raw input. The self-attention mechanism is then applied, allowing the model to focus on different parts of the input that are most relevant for estimating the noise. To improve accuracy, a mask is used to guide the model's attention, ensuring that it focuses only on the noisy parts of embeddings for better error correction. The output of the decoding process $\hat{\boldsymbol{w}}$ represents the model’s prediction of the multiplicative noise in the received codeword. This vector identifies and flips the bits at positions with detected errors in received vector \( \boldsymbol{r} \). The last step is to obtain the corrected codeword $\hat{\boldsymbol{c}}$ as a hard-decision on the vector $\boldsymbol{w} \odot (2\boldsymbol{r}_\text{hd}-1)$, where $\odot$ is an element-wise multiplication.

\setcounter{example}{2}
\begin{example}[ECCT for Hamming~$(7,4)$ (Cont.)]
Let the predicted multiplicative noise be the vector \( \hat{\boldsymbol{w}} = [5.2, 5.7, 2.4, -6.1, 6.1, 5.5, 6.8] \). The BPSK-modulated version of $\boldsymbol{r}_\text{hd}$ is given by $[-1,+1,-1,+1,+1,+1,-1]$. The resulting vector is $\boldsymbol{w} \odot (2\boldsymbol{r}_\text{hd}-1) = [-5.2, 5.7, -2.4, -6.1, 6.1, 5.5, -6.8]$. A hard-decision on this vector gives \( \hat{\boldsymbol{c}} = [0, 1, 0, 0, 1, 1, 0] \), where the negative value in the fourth position of the vector $\hat{\boldsymbol{w}}$ is responsible for the bit flip that gives a correctly codeword.
\end{example}

CrossMPT, built on a cross-attention mechanism, represents the state-of-the-art in error correction transformers \cite{park2024crossmptcrossattentionmessagepassingtransformer}.
Similar to self-attention, cross-attention uses two distinct inputs but produces one single output. In CrossMPT, the magnitude $\boldsymbol{|r|}$ and the sign of the syndrome  \( \boldsymbol{s}(\boldsymbol{r}) \) are separately embedded and processed through cross-attention decoding, where two self-attention models iteratively exchange information, similar to BP. The output and post-processing are similar to ECCT, where output combines with the received codeword $\boldsymbol{r}$ to correct errors. More details can be found in \cite[Sec.~4]{park2024crossmptcrossattentionmessagepassingtransformer}.

\subsection{Numerical Results}
Here we present numerical results for different BCH codes as examples of short and medium-length codes, where ECCT and CrossMPT are expected to perform well. We consider BCH~$(31,16)$, BCH~$(63,51)$, and BCH~$(127,64)$ codes. The performance of ECCT and CrossMPT is compared against hard-decision bounded distance decoding (BDD) and OSD with two different values of $q$, which determine the number of test patterns (see Sec.~\ref{OSD}). We consider the two simplest cases of OSD: $q=0$ (1 test pattern), and $q=1$ ($k$ test patterns). We use OSD as a comparison for ECCT and CrossMPT as OSD is a soft-input decoding algorithm, just like ECCT and CrossMPT. Results are shown for $d=128$, $N=6$ and $N=10$, where $d$ denotes the embedding dimension, and $N$ decoding iterations.

Fig.~\ref{fig:ECCT results} shows the BER results obtained. For the BCH~$(31,16)$ code (left), ECCT and CrossMPT offer large gains with respect to HD decoding, with performances close to ML (dashed blue line). CrossMPT shows slightly better performance than ECCT. Although ECCT and CrossMPT outperform OSD with $q=0$, OSD with $q=1$ offers near-ML performance, providing better performance than ECCT and CrossMPT. The results for the BCH $(63,51)$ code (middle) are very similar and allow us to conclude that for these two BCH codes, a simple off-the-shelf OSD decoder ($q=1$, with $16$ or $51$ test patterns) outperforms ECCT and CrossMPT. It is important to mention that training ECCT for these two BCH codes requires optimization of about $1.2$ million parameters, and for CrossMPT, the number is almost the same.

We conclude by showing results for the BCH~$(127,64)$ code. The results in Fig.~\ref{fig:ECCT results} (right) show that ECCT with $N=6$ is not competitive, even with respect to hard-decision decoding. We also report results for ECCT with $N=10$, as well as CrossMPT with $N=6$ and $N=10$, which show slight improvements with respect to ECCT with $N=6$. All three approaches exhibit nearly identical performance, but still far worse than HD decoding. Again, OSD with $q=1$, whose main computation complexity is the Gaussian elimination required for finding the most reliable basis, offers excellent performance (in this case with $64$ test patterns), about $2$~dB better than transformer-based decoder.

\section{Conclusions}\label{conclusion}
We provided a comprehensive analysis of four neural network-based decoders for error correcting codes. SLNN and MLNN decoders were shown to always achieve maximum likelihood performance without requiring training. These decoders are limited to be used with short codes only, as both SLNN and MLNN always require at least one layer with an exponentially increasing numbers of neurons (number of codewords). Transformer-based approaches were shown not to be competitive when compared to traditional decoders such as ordered statistics decoding. The results presented in this paper raise important concerns about the viability of neural network-based decoders, particularly in short and medium blocklength regimes. Future work includes a precise complexity comparison between NN-based and traditional decoders.

\bibliographystyle{IEEEtran}
\bibliography{References}

\begin{thebibliography}{10}
\providecommand{\url}[1]{#1}
\csname url@samestyle\endcsname
\providecommand{\newblock}{\relax}
\providecommand{\bibinfo}[2]{#2}
\providecommand{\BIBentrySTDinterwordspacing}{\spaceskip=0pt\relax}
\providecommand{\BIBentryALTinterwordstretchfactor}{4}
\providecommand{\BIBentryALTinterwordspacing}{\spaceskip=\fontdimen2\font plus
\BIBentryALTinterwordstretchfactor\fontdimen3\font minus
  \fontdimen4\font\relax}
\providecommand{\BIBforeignlanguage}[2]{{%
\expandafter\ifx\csname l@#1\endcsname\relax
\typeout{** WARNING: IEEEtran.bst: No hyphenation pattern has been}%
\typeout{** loaded for the language `#1'. Using the pattern for}%
\typeout{** the default language instead.}%
\else
\language=\csname l@#1\endcsname
\fi
#2}}
\providecommand{\BIBdecl}{\relax}
\BIBdecl

\bibitem{intro_surveyfec}
G.~Tzimpragos, C.~Kachris, I.~B. Djordjevic, M.~Cvijetic, D.~Soudris, and
  I.~Tomkos, ``A survey on {FEC} codes for 100{G} and beyond optical
  networks,'' \emph{IEEE Commun. Surveys Tuts.}, vol.~18, no.~1, pp. 209--221,
  Oct. 2016.

\bibitem{ml_problem}
E.~Berlekamp, R.~McEliece, and H.~van Tilborg, ``On the inherent intractability
  of certain coding problems (corresp.),'' \emph{IEEE Trans. Inf. Theory},
  vol.~24, no.~3, pp. 384--386, 1978.

\bibitem{bitflip}
R.~Gallager, ``Low-density parity-check codes,'' \emph{IRE Trans. Inf. Theory},
  vol.~8, no.~1, pp. 21--28, 1962.

\bibitem{1363033}
D.~Hocevar, ``A reduced complexity decoder architecture via layered decoding of
  {LDPC} codes,'' in \emph{Proc. IEEE Workshop Signal Process. Syst. (SIPS)},
  Austin, TX, USA, Oct. 2004.

\bibitem{fossorier}
M.~Fossorier and S.~Lin, ``Soft-decision decoding of linear block codes based
  on ordered statistics,'' \emph{IEEE Trans. Inf. Theory}, vol.~41, no.~5, pp.
  1379--1396, 1995.

\bibitem{dlbasedchanneldecoding}
T.~Gruber, S.~Cammerer, J.~Hoydis, and S.~t. Brink, ``On deep learning-based
  channel decoding,'' in \emph{Proc. Conf. Inf. Sci. Syst. (CISS)}, Baltimore,
  MD, USA, Mar. 2017.

\bibitem{SLNN1}
A.~Di~Stefano, O.~Mirabella, G.~Di~Cataldo, and G.~Palumbo, ``On the use of
  neural networks for {Hamming} coding,'' in \emph{Proc. IEEE Int. Symp.
  Circuits Syst. (ISCAS)}, Singapore, June 1991.

\bibitem{SLNN2}
C.~T. Leung, R.~V. Bhat, and M.~Motani, ``Low-latency neural decoders for
  linear and non-linear block codes,'' in \emph{Proc. IEEE Global Commun. Conf.
  (GLOBECOM)}, Waikoloa, HI, USA, Dec. 2019.

\bibitem{SLNN3}
------, ``Low latency energy-efficient neural decoders for block codes,''
  \emph{IEEE Trans. Green Commun. Netw.}, vol.~7, no.~2, pp. 680--691, 2023.

\bibitem{MLNN1}
------, ``Multi-label neural decoders for block codes,'' in \emph{Proc. Int.
  Conf. Commun. (ICC)}, virtual event, June 2020.

\bibitem{MLNN2}
V.~Malik, R.~Ghosh, and M.~Motani, ``Achieving low complexity neural decoders
  via iterative pruning,'' in \emph{Proc. Cof. Neural Inf. Process. Syst.
  (NeurIPS)}, New Orleans, LA, USA, Dec. 2022.

\bibitem{MLNN3}
Y.~Lei, M.~He, H.~Song, X.~Teng, Z.~Hu, P.~Pan, and H.~Wang, ``A
  deep-neural-network-based decoding scheme in wireless communication
  systems,'' \emph{Electronics}, vol.~12, no. 13: 2973, July 2023.

\bibitem{MLNN4}
S.-W. Wang, C.-M. Huang, C.-C. Yang, and G.-Y. Yang, ``Implementation of neural
  network-based linear block code decoder in {SAC-OCDMA} system,'' in
  \emph{Proc. IEEE VTS Asia Pacific Wireless Commun. Symp. (APWCS)}, Tainan,
  Taiwan, Aug. 2023.

\bibitem{b9}
S.~E. El-Khamy, E.~A. Youssef, and H.~M. Abdou, ``Soft decision decoding of
  block codes using artificial neural network,'' in \emph{Proc. IEEE Symp.
  Comput. Commun. (ISCC)}, Alexandria, Egypt, June 1995.

\bibitem{dl-survey}
\BIBentryALTinterwordspacing
T.~Matsumine and H.~Ochiai, ``Recent advances in deep learning for channel
  coding: A survey,'' June 2024. [Online]. Available:
  \url{https://arxiv.org/abs/2406.19664}
\BIBentrySTDinterwordspacing

\bibitem{ECCT}
Y.~Choukroun and L.~Wolf, ``Error correction code transformer,'' in \emph{Proc.
  Cof. Neural Inf. Process. Syst. (NeurIPS)}, New Orleans, LA, USA, Dec. 2022.

\bibitem{park2024crossmptcrossattentionmessagepassingtransformer}
\BIBentryALTinterwordspacing
S.-J. Park, H.-Y. Kwak, S.-H. Kim, Y.~Kim, and J.-S. No, ``{CrossMPT}:
  Cross-attention message-passing transformer for error correcting codes,'' May
  2024. [Online]. Available: \url{http://arxiv.org/abs/2405.01033}
\BIBentrySTDinterwordspacing

\bibitem{nbp1}
E.~Nachmani, E.~Marciano, L.~Lugosch, W.~J. Gross, D.~Burshtein, and
  Y.~Be’ery, ``Deep learning methods for improved decoding of linear codes,''
  \emph{IEEE J. Sel. Topics Signal Process.}, vol.~12, no.~1, pp. 119--131,
  Feb. 2018.

\bibitem{vaswani2023attentionneed}
A.~Vaswani, N.~Shazeer, N.~Parmar, J.~Uszkoreit, L.~Jones, A.~N. Gomez,
  L.~Kaiser, and I.~Polosukhin, ``Attention is all you need,'' in \emph{Proc.
  Cof. Neural Inf. Process. Syst. (NeurIPS)}, Long Beach, CA, USA, Dec. 2023.

\end{thebibliography}

\end{document}